\documentclass[10pt,journal]{IEEEtran}
\usepackage{times,amsmath,epsfig}
\title{Maintaining Virtual Areas on FPGAs using Strip~Packing with Delays}
\author{%
{%
Josef Angermeier{\small $^{\#1}$},
S\'andor~P.~Fekete{\small $^{*}$},
Tom~Kamphans{\small $^{*2}$},
Nils~Schweer{\small $^{*}$},
J\"urgen Teich{\small $^{\#}$}%
}
\vspace{1.6mm}\\
\fontsize{10}{10}\selectfont\itshape
$^{\#}$\,%
Department of Computer Science 12, University of Erlangen-Nuremberg\\
Erlangen, Germany\\
\fontsize{9}{9}\selectfont\ttfamily\upshape
\{angermeier, teich\}@cs.fau.de%
\vspace{1.2mm}\\
\fontsize{10}{10}\selectfont\rmfamily\itshape
$^{*}$\,%
Department of Computer Science, Braunschweig University of Technology\\
Braunschweig, Germany\\
\fontsize{9}{9}\selectfont\ttfamily\upshape
\{s.fekete, n.schweer\}@tu-bs.de, tom@kamphans.de%
}
\begin{document}
\maketitle
\begin{abstract}
  The computing resources available on dynamically partially
  reconfigurable devices increase every year enormously. In the near
  future, we expect that many applications run on a single
  reconfigurable device. In this paper, we present a concept for
  multitasking on dynamically partially reconfigurable systems called
  {\em virtual area management}. We explain its advantages, show its
  challenges, and discuss possible solutions. Furthermore, we investigate
  one problem in more detail: Packing modules with time-varying 
  resource requests.
  This problem
  from the reconfigurable computing field results in a completely new
  optimization problem not tackled before. ILP-based and heuristic
  approaches are compared in an experimental study and the drawbacks
  and benefits discussed.

\end{abstract}

\footnotetext[1]{Supported by DFG grant TE~163/14-2, 14-3,
project ``ReCoNodes'', as part of the Priority Programme 1148.
``Reconfigurable Computing''.}
\footnotetext[2]{Supported by DFG grant FE~407/8-2, 8-3, project ``ReCoNodes'',
as part of the Priority Programme 1148, ``Reconfigurable Computing''.}

\section{Introduction}
Reconfigurable devices offer more and more space and functionality
over time and will probably continue to do so in the future. Yet, huge hardware
applications can already be instantiated on the reconfigurable chips,
or even arrays of processors. Furthermore, nowadays reconfigurable
chips are mostly used to instantiate a single application with
multiple modules, which might not all be necessary at each instant in
time. But as reconfigurable devices evolve further, the offered
resources will be large enough to also run completely different
applications simultaneously. This development also took place
in the software world: In the beginning of the nineties, most personal
computers used operating systems such as MS-DOS, 
which allowed just one single
software application to be executed, alone with some drivers. 
Since then, the increased computing resources have allowed us to run multiple
applications simultaneously.

Similar to the software world, multitasking will lead reconfigurable
devices to higher efficiency, but will also raise several challenges
that must be solved. For example, in order to provide reliability and
security, we must make sure that no application can violate the
execution of another one. In software this is solved by running each
application in its own virtual address space. Each application knows
only the virtual addresses of its own resources and parts of the
operating system. The virtual addresses are translated at runtime into
physical addresses in the memory. Thus, pages of different
applications may lie physically next to each other, but only the
approved application may access them. Furthermore, the concept of the
virtual address space also facilitated writing software, as each
application did not need to bother about the positions of other
applications but can assume to be the only user of the processor and
the memory.

In order to work out concepts for allowing multitasking on
reconfigurable devices, we oriented ourselves on concepts of the
software world such as virtual address space. But, as there are some
major differences between hardware and software, one cannot just put
one concept unchanged from software to reconfigurable devices. One
important difference is that on reconfigurable devices, different
modules may be executed in parallel while software usually works more
sequentially.

The basic idea of our concept which we called ''Virtual Area
Management'' is to partition the available FPGA area into different
regions. Each hardware application obtains one contiguous region,
which may grow or shrink depending on the applications resource
needs. Each running hardware application can reconfigure hardware
modules in its region and free or try to allocate more area. In this
process, each application does not know the physical position of its
modules, but just the relative positions. Thus, it cannot reconfigure
any region that belongs to a different application and is concerned
only with its own resources. Intermodule communication takes place
only in the region of one application. Thus, the modules that must
communicate with each other are automatically grouped by position to
each other.

Communication between the partial modules and with the input and
output periphery is an important point. We assume that the application
modules are provided with some means to communicate with each other
and to the FPGA's I/O-ports independently from their position (e.g.,
as described in \cite{koch_fccm}). 
There are several possible ways to achieve this
goal. We solved the communication problems by using our self-developed
platform called ESM (see
\cite{AGMTFV_ESM_2007},\cite{bobda_erlangen_2005}). It 
offers---amongst others---a
so called crossbar device, which dynamically routes the
input and output signals to the position of the corresponding
module. 
Moreover, the modules can
communicate to each other using the crossbar.
Thus, we can assume that the
modules can be placed independent from the positions
of the I/O periphery and other the positions of other modules. 

\subsection{Related Work}
\subsubsection{Reconfigurable Computing}
Brebner~\cite{brebner_virtual_1996,brebner_swappable_1997}
addressed the problems involved in presenting to a software-oriented user 
a larger virtual hardware resource that is
implemented using smaller physical FPGA hardware. Their
approach is based on using swappable units, and a prototype operating
system is described that demonstrates operational steps. In contrast
to that work, this paper does not address the problem of overcoming the
physical constraints given by a small FPGA, but tackles the problem,
how to run multiple applications on a single reconfigurable resource.

Bazargan et~al.~\cite{bazargan_fast_2000} present fast online
placement methods for dynamically reconfigurable systems, as well as
offline 3D placement algorithms. Hereby, partial modules are to be
placed completely independent of each other, and the inter-module
communication problem is not addressed. Steiger et 
al.~\cite{steiger_online_2003} and Diessel et 
al.~\cite{diessel_dynamic_2000} further improve scheduling methods for
partially reconfigurable systems. Our approach is based on the
differentiation between applications and modules. Modules belonging to
the same application are placed nearby, such that inter-module
communication can be as efficient as possible.  Different scheduling
subproblems have been addressed meanwhile; for example, 
scheduling with respect to the reconfiguration 
overhead~\cite{resano_hybrid_2005}.
Banerjee et al.~\cite{banerjee_physically-aware_2005}
take into account hardware-software partitioning decisions for a
fast execution of an application. But in contrast to our paper, all
these approaches still focus on executing a single application.

Some operating systems for reconfigurable embedded platforms were developed
\cite{steiger_operating_2004,wigley_development_2001,walder_reconfigurable_2003}. Such an operating
system provides a minimal programming model and a runtime system. The
runtime system performs online task and resource
management. Scheduling problems are formulated for the 1D and 2D
resource models and developed heuristics are compared to each
other. Resources of the operating system and the user applications are
clearly differentiated, but that is not the case for the resource
access of multiple applications running on the FPGA. Our
paper suggests a compromise for the inter-module communication
problem: Modules belonging to the same application should placed 
nearby to each other, such that they can exchange data
efficiently. Modules belonging to different applications are not
necessarily placed nearby. Additionally, in contrast to our application model,
no application can shrink and grow during runtime. The
focus in former works was put more on hardware and software
abstraction, here it is on securely running multiple, dynamic
hardware applications.

Many recent works focused on achieving optimal performance to put
multiple tasks onto one FPGA within this context.
Cordone et al.~\cite{crrss-pstgp-09} and
Redaelli et al.~\cite{rss-tscpa-08} specified a new model for
partitioning and scheduling on partially dynamically reconfigurable
hardware. The different applications are represented by a
task graph and the aim is to obtain a total execution time near
optimality by taking reconfiguration and communication times into
account. However, this approach does not allow dynamic behavior, such
that according to the current state of the resources, the tasks can
select on their own which module implementation to reconfigure next
and at which nearby position to place it.

The approach by Cardoso~\cite{c-ctpsf-03}  also considers the topic of
resource  virtualization on  FPGA devices,  achievable due  to dynamic
reconfiguration  capabilities.  Hereby,  a new  temporal  partitioning
algorithm    is     proposed.    The    model     by    Banerjee    et
al.~\cite{bbd-ipchs-06}  furthermore also supports  HW/SW partitioning
of  the  tasks.  However,  both  approaches  are  also  based  on  the
assumption  that the  running  times  of each  task  can be  estimated
roughly, and  that the worst-case  execution time does not  differ too
much from the average case. In contrast to that, our approach may also
be engaged for the online case, where this must not be the case.

\subsubsection{Packing}

It turns out that placing hardware modules with growing and shrinking
area resources amounts to strip packing. The classical strip packing
problem was first considered by Baker et al.~\cite{bcr-optd-80}. In
this problem a set of rectangles must be packed into a strip of
semi-infinite height and width 1 such that the total height of the
packing is minimized. They showed that in the online case the bottom
left heuristic does not guarantee a constant competitive
ratio for packing a sequence of rectangles.  For the offline case they proved
an upper bound of $3$ for a sequence of rectangles and of $2$ on the
competitive ratio for a sequence of squares; both analyses require the
elements to be sorted. Later Kenyon and R{\'e}mila designed a fully
polynomial time approximation scheme for the offline
setting~\cite{kr-asp-96}. For the online case Baker and
Schwarz~\cite{BakSch83} introduced the so-called shelf algorithms with
an competitive ratio that can be made arbitrarily close to 1.7. Csirik
and Woeginger~\cite{cw-saosp-97} showed a lower bound of $1.69103$ for
any shelf algorithm and introduced an algorithm whose asymptotic
worst-case ratio comes arbitrarily close to this value.

In the classical game of Tetris the aim is to find online placements
for a sequence of objects---not all having rectangular shape---such
that space is utilized as well as possible. In this
process, no item can ever move upward, no collisions between objects
must occur, an item will come to a stop if and only if it is supported
from below, and each placement must be placed before the next item
arrives. Obviously, there is a
slight difference in the objective function, as Tetris aims at filling
rows. In actual optimization scenarios, this is less interesting, as
it is not critical whether a row is used to precisely 100\%. 
Even when disregarding the difficulty of ever-increasing
speed, Tetris is notoriously difficult: As shown by Breukelaar et
al.~\cite{bdhhkl-tihea-04}, Tetris is PSPACE-hard, even for the
original, limited set of different objects. Azar and
Epstein~\cite{ae-tdp-97} considered Tetris-like online packing of
rectangles into a strip where each item must be moved on a collision-free 
path to its final position which does not have to supported from
below. Just like in Tetris, they considered the situation with or
without rotation of objects. For the case without rotation, they
showed that no constant competitive ratio is possible, unless there is
a fixed-size lower bound of $\varepsilon$ on the side length of the
objects, in which case there is an upper bound of
$O(\log\frac{1}{\varepsilon})$. For the case in which rotation is
possible, they showed a 4-competitive strategy, based on shelf-packing
methods, with all rectangles being rotated to be placed on their
narrow sides. Coffmann, Downey, and Winkler~\cite{cdw-prs-02}
considered probabilistic aspects of online rectangle packing with
Tetris constraint, without allowing rotations. If rectangle side
lengths are chosen uniformly at random from the interval $[0,1]$, they
showed that there is a lower bound of $(0.31382733...)n$ on the
expected height of the strip. Using another kind of level-type
strategy, which arises from the bin-packing--inspired {\em Next Fit
  Level}, they established an upper bound of $(0.36976421...)n$ on the
expected height. 
Fekete et al.~\cite{fks-olsp-09p} considered an Tetris-like online packing
with gravity; that is, every item must be supported from
below in its final position. For squares they gave an algorithm with
competitive ratio 2.6154. 

Note that none of the previous works allows
stretching the objects in any direction.

\section{Virtual Area Management}

\subsection{Main Idea}

\begin{figure}[t]
        \centering \includegraphics[width=8.0cm]{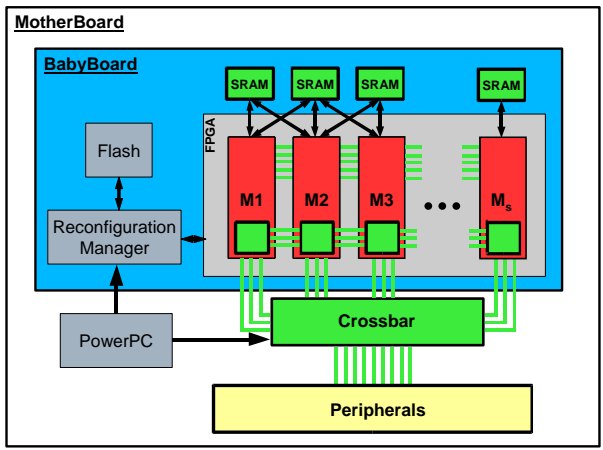}
        \caption{Architecture overview our ESM platform (see \cite{AGMTFV_ESM_2007},\cite{bobda_erlangen_2005} for more details).} \label{fig:esm-arch}
\end{figure}

Our concept is aimed at partially dynamically reconfigurable
architectures, where modules loaded onto the reconfigurable device may
be exchanged at runtime.  A typical structure of such a device is
given in Fig.~\ref{fig:esm-arch}.  Yet devices are available which
allow the reconfiguration of columns only, while newer platforms also
allow reconfiguration of certain contiguous cells. The former is
called 1D reconfiguration, the latter 2D reconfiguration. Our concepts
are applicable to both architectures. Furthermore, the platform should
consist of one control CPU, which may be placed externally or be
included into the reconfigurable device.

\begin{figure}[t]
        \centering \includegraphics[width=7.5cm]{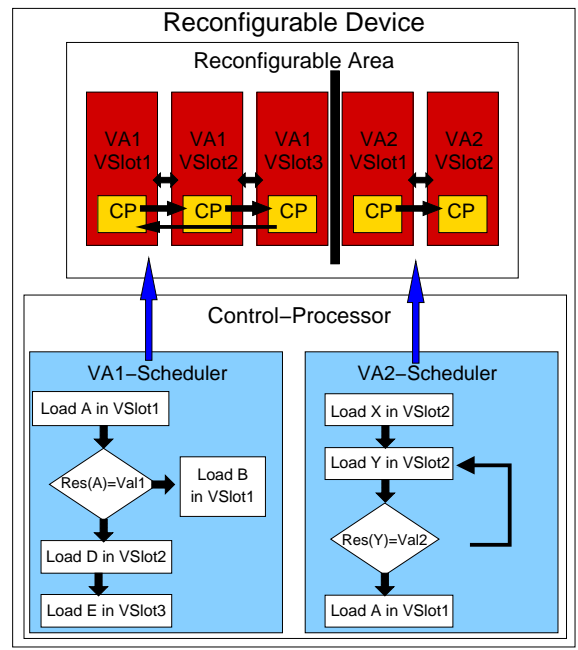}
        \caption{Two different hardware applications running on a 1D
          partially dynamically reconfigurable device: The first three
          slots form the first virtual area (VA1), the remaining the
          second virtual area (VA2). Modules in one VA can communicate
          with each other, e.g. by using their communication point
          (CP) belonging to reconfigurable bus. Neither the first nor
          the second application knows the exact position of its VA on
          the reconfigurable device; only the size of the available
          area and the relative positions of the reconfigurable
          modules loaded in its virtual area are
          known.} \label{fig:esm_with_vesms}
\end{figure}

To run different applications on the reconfigurable device,
we propose to partition the available reconfigurable area into so
called {\em virtual areas} (VAs). For performance reasons, the different VAs
should consist of contiguous reconfigurable area units. 
As the required amount of resources of an
application changes over time, the size of the
virtual area may change dynamically over the running time of its
application. The mapping of virtual area to physical reconfigurable
area is done by the control processor. Each hardware application being
executed on the reconfigurable device has its own software control
thread running on this CPU, see Fig.~\ref{fig:esm_with_vesms}. These
threads request the initially required area and transmit
changes in the requirements.
An operating system
service on the software side takes the requests and is in charge of
the management of the virtual areas. This secure operating system unit
maps the virtual area units to the corresponding
physical dynamically reconfigurable area units. The virtual area
management unit can be compared to the memory management unit (MMU) in
the software world: both handle the translation of virtual to physical
memory positions. Furthermore, the corresponding application software
threads do not know the actual physical positions of the reconfigured
modules, only the relative positions of each reconfigurable module
to each other. Each application simply requests to load a specified
module to a virtual position in the assigned virtual area. See the
example in Fig.~\ref{fig:esm_with_vesms}: The second
application with its virtual area VA2 issues a request to load a
specified bit-file called ``X'' to the virtual address (here called:
{\em VSlot}) `2'. It does not know that this virtual address corresponds
physically to the last reconfigurable unit on the reconfigurable
device. It knows only that the module loaded there is on the right
side of a module loaded to the virtual address (VSlot) `1'. As each
application is allowed to specify only a virtual address corresponding
to its virtual area in which to place the module, it cannot place a module
into an area belonging to a different application.

The concept is transparent to the applied intermodule communication in
each virtual area: Each application can choose its preferred
communication method (e.g., bus system, neighbor to neighbor
communication over bus macros). When only contiguous virtual
areas are used, the communicating modules are grouped together
automatically and communication overhead is minimized. Furthermore,
communication between different applications or heterogeneities of the
reconfigurable area is supported easily by extending the operating
system.

Yet there are many other approaches to schedule tasks of different
applications, but in contrast to them our approach of virtual areas
combines a localization strategy, putting tasks which belong to one
application together, and highly dynamic applications, which can
individually take different module selection and placement decision
based on the current context.

\subsection{Advantages}

The idea of virtual area management offers the following advantages
when running multiple applications on a single dynamically partially
reconfigurable device.

\begin{itemize}

\item \emph{Resource accounting}: The concept can be easily used to
  prevent applications from using too many resources at runtime. 
  One might want to restrict the resources
  granted to an application for each execution or just in a
  certain context in order. The goal is, for example,
  to reduce the power consumption
  or to accelerate more important applications in a system where
  certain applications have lower priorities than others.
  The concept of virtual area management provides a
  limitation mechanism by granting just a certain amount of, for example,
  reconfigurable area to an application.

\item \emph{Resource protection}: Each application controls only the
  reconfiguration within its virtual area, as it can only specify those
  virtual positions that are valid in its own VA. The virtual
  area management unit checks for each request if
  the virtual address is valid and translates it into a physical
  position on the reconfigurable device. This way, no application can
  load any module to the area of another application and the
  applications are protected. Thus, an error in one application cannot
  harm the execution of all other applications and lead to a complete
  system breakdown.

\item \emph{Support for differing scheduling and placement
    procedures}: Different applications require different scheduling
  and placement methods to increase performance. Instead of designing
  complex scheduling algorithms that try to meet all the
  requirements (e.g., periodic or aperiodic, with or without
  deadlines) by introducing various priority classes, each
  application gets its own virtual area and specifies the
  best scheduling strategy. This may include a simpler and faster
  implementation of new hardware applications within an existing
  system.

\item \emph{Area position transparency and programming dynamic
    applications}: The absolute positions' independence of the
  executed modules, in the following called {\em area position
  transparency}, offers a new programming model. It
  allows to write dynamic applications, which subject to
  the current resource context decide on their own how to
  proceed. 
  Depending on the assigned area, an application can either use an
  implementation that offers more performance but needs more
  area, or another one that takes longer but uses less area. 
  Another new possibility
  is that the application can decide how to increase and shrink
  depending on the current area usage context. An application can ask
  the virtual area management unit, if it can grow to the left or
  right, or at the bottom, and, depending where some unused area is
  available, it can decide to put a partial module there and
  instantiate some appropriate communication module to transfer data
  there and back. Thus, at each run, an application can operate differently in
  its amount of resource usage. Before, trade-off decisions where also
  possible for a single application, but the new idea is to let
  each application decide in its own control program its next steps
  depending on the behavior of the other applications.

\end{itemize}

\subsection{Challenges}

First, there are some technical problems to be solved. For area
position transparency, the placement of a module should not
be limited to one single position. Furthermore, there might be
heterogeneities on the reconfigurable device which require
different implementation for different positions. A possible solution to this
problem is to generate implementations for the module for each
possible position where the corresponding virtual area can be
placed. The corresponding module implementation bit-files can be
compressed to save some space. Another option is to apply a
single generated module bit-file and relocate this file; that is, adapt it to
the corresponding position. We solved this technical
issue in the following way: Our experimental board is equipped with a
reconfiguration manager device that manipulates the corresponding
bit-files before the reconfiguration of the corresponding device.

A further technical challenge lies in the communication of the partial
modules to the external periphery (e.g., video, audio) over the
input/output pins at the border of the FPGA.
Our experimental board has a
crossbar device that routes I/O signals dynamically from the
periphery to the current position of the partial module.

A larger challenge is to fulfill the changing resource requirements of
the different applications. Every application has a different resource
usage profile which depends on the inputs specified. An application
called with a larger problem instance to solve needs also more
resources. Additionally, the resource usage profile also depends on
the execution context. The application may behave differently
depending on how many area resources are available at each position.
Furthermore, the resource requirement depending on the
inputs of an application may or may not be known in advance (or at least can be
estimated, e.g., in a numerical application based on the required
precision of the solutions). The former is called the offline case,
the latter the online case.

\begin{figure}[t]
        \centering ~\qquad\includegraphics[width=8.5cm]{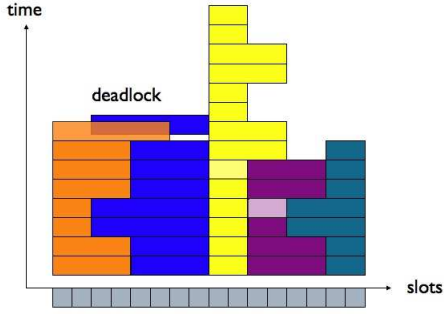}
        \caption{Both the first and the second application (from the
          left), request more area resources in order to proceed. A
          deadlock occurs, because no more resources are available to
          meet any request.\newline~} \label{fig:deadlock}
\end{figure}

In the online case, a deadlock is possible when two applications currently
running on the reconfigurable device can only proceed in their task
when an resource increase is granted, see
Fig.~\ref{fig:deadlock}. The blocks represent the occupied area of one
application at different points in time. The leftmost application 
needs two more area units, as does the application second from the
left. Furthermore, no application can give up its accumulated
resources, as saving and restoring states is widely considered too
expensive on FPGAs. Such a deadlock must be prevented. 
A commonly used solution is to allow hardware task preemption: If not
enough resources are available to meet increased resource demands, the
state of one application is stored in external memory. At a
later point in time, the application is loaded again on the
reconfigurable device and the state be restored again to continue its
processing. 
Another approach is not to wait until a deadlock scenario
actually happens, but to check beforehand that it cannot get this
far: Assuming that the maximal size of a request is known,
an area shared between two applications is
granted exclusively to just one application, but not one part of it to
the one application, and another part to the other application.

\begin{figure}[b]
        \centering \includegraphics[width=8.5cm]{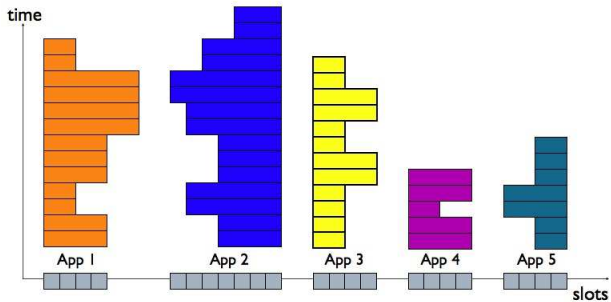}
        \caption{Five applications modules, each with different resource 
          demands over time for their specific inputs.} \label{fig:profile}
\end{figure}

In the following, we consider the offline problem: The resource
usage profiles for specific inputs of a series of applications is
known a priori, or can be estimated roughly. 
An example of application resource profiles is given
in Fig.~\ref{fig:profile}. 
Note that in the offline case deadlocks cannot occur:
We search for feasible solutions (i.e., 
schedules where no resource requests overlap) only.

\begin{figure}[t]
        \centering 
        \includegraphics[width=8.5cm]{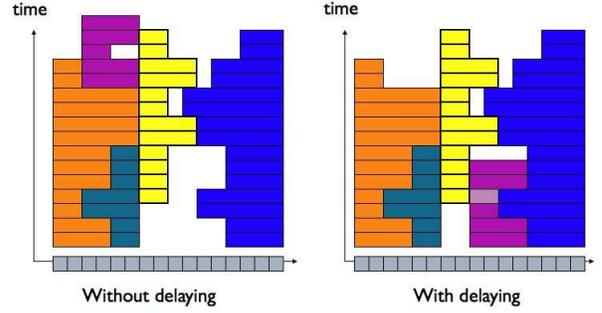}
        \caption{Left: Schedule of the resource demands on the
          reconfigurable device under the assumption that running
          applications cannot wait for a resource grant.
          Right: Schedule for the same demands, but with the
          option to delay resource grants.
        } \label{fig:example_with_wait}
\end{figure}

Hardware task preemption can be used to increase the area
usage. However, saving and restoring the states of the hardware
applications can be very costly in time and memory. 
An approach that balances 
reconfiguration costs and efficient resource usage is 
to allow that requests may be delayed by the scheduler. 
The application keeps its currently occupied area, but remains
idle until the request is fulfilled.
Compare the two schedules for our example 
shown in Fig.~\ref{fig:example_with_wait}: The schedule shown on the left
hand side is a solution for scheduling the application modules without
delaying requests. On the right hand side, we allow that requests are
postponed. Using this option for the forth request of the forth
module, we achieve a better makespan. 

\section{Packing Application Modules}
We consider the problem {\em FPGATris}: Scheduling modules whose 
resource requests (i.e., space on an FPGA) varies over time. 
This may be, for example, a
router module that needs more resources if the traffic increases.
We assume that a module
occupies a certain number of slots on the FPGA, but
requests only complete slots. Thus, we model
the FPGA as a one-dimensional array.
Furthermore, we assume that time is discrete; that is, 
requests are multiples of a fixed-size time slot.
Now, scheduling a sequence of modules with time-varying resources 
corresponds to strip packing: The width of the strip is
the number of slots on the FPGA; the height corresponds to the
time axis. Thus, we use height and time synonymously.
Moreover, we assume that every module occupies a 
base slot and extends to the left or
to the right of the base slot.%
\footnote{For convenience, we consider only the case of growing
either to the left or to the right of the base slot. The generalization
to both sides is straightforward.}.

\begin{figure}[b]
\centerline{\input{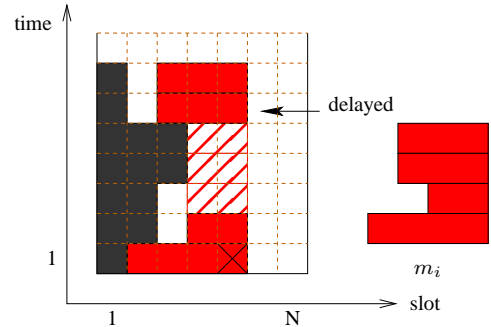}}
\caption{We can place the module $m_i$ at the position marked by
$\times$ (the {\em base slot} of $m_i$), 
if we {\em delay} the third request. That is, the second
request stays on the FPGA until the third request can be fulfilled.
\label{fpgatris-delay-fig}}

\end{figure}

We are allowed to {\em delay} a request; that is, 
we may stretch the modules along the time axis; see 
Fig.~\ref{fpgatris-delay-fig}.
Our goal is to minimize the {\em makespan} (i.e., the time needed to fulfill
all requests). 
For the strip-packing problem, this goal corresponds to minimizing 
the height of the occupied part of the strip.

\subsection{An ILP}
We are given a strip of width $N$ 
(e.g., an FPGA with $N$ slots and a time axis) and
want to place $M$ modules.
Each module, $m_i$, is given as a sequence
of {\em requests}.
Let $\ell_i$ denote the length of the request sequence for module $m_i$ and
$(i,j)$ the $j$th request of $m_i$. The size of $(i,j)$ is given by
$r_{ij} \!\in\! \mathrm{Z\kern-.33em Z}\backslash \{0\}, 
1\!\leq\! j\!\leq\! \ell_i$. 
For $r_{ij} > 0$ the module expands to the right of the base slot, 
for $r_{ij} < 0$ to the left.

\smallskip\noindent
For the ILP, we introduce four kinds of variables:
\begin{itemize}
\item the {\em slot assignment variables}, $x_{si}$
\item the {\em time assignment variables}, $y_{tij}$
\item the {\em occupancy variables}, $z_{stij}$
\item the {\em usage variables}, $u_t$
\end{itemize}

The first two types specify when and where a request
is scheduled. More precisely, setting $x_{si}$ to 1 indicates 
that module $m_i$ is scheduled in slot $s$. 
Setting $y_{tij}$ to 1 indicates that 
request $(i,j)$ of module $m_i$ is scheduled at time $t$. That is,
module $m_i$ occupies the following slots at time $t$:
$$s, \ldots, s+r_{ij}-1 \mbox{ \rm for }
r_{ij} > 0,$$
$$s+r_{ij}+1, \ldots, s \mbox{ \rm for }
r_{ij} < 0,$$

\noindent
where $s$ is the base slot of module $m_i$. 
For every 
$i$ there is exactly one $x_{si}$ with $x_{si}=1$ and
for every 
$(i,j)$ there is exactly one $y_{tij}$ with $y_{tij}=1$.

Usually (i.e., if $|r_{ij}| > 1$) a request occupies more than one slot when
executed. Moreover, if the request $(i,j+1)$ is delayed, $(i,j)$ 
remains on the FPGA for more than one time unit
(i.e., it occupies more the one time row).
To keep track of the occupied slots, we set $z_{stij}$ to 1, if 
slot $s$ is occupied by request $(i,j)$ at time $t$.
The usage variables simply specify which
time steps are used. 

Clearly, the FPGA's size, $N$, and the number of modules, $M$, 
strongly determine the size of an ILP and, in turn, the time needed to solve it.
In addition, we assume that an upper bound, $T$, on the number of time steps
is given. The closer this bound is to the optimum, the smaller the 
resulting ILP. This upper bound can be obtained, for example, using
the tabu search in Sect.~\ref{tabu-sect}.\\

\subsubsection{Constraints}
\paragraph{Assignment Constraints} 
Each request must be scheduled exactly once. That is,
for every $(i,j)$ we have to set exactly one $x_{si}$ to 1 (to assign
a slot for $(i,j)$) and exactly one $y_{tij}$ (to assign a start time
for $(i,j)$).
The following constraints express these conditions:
\begin{equation}\label{fpgatris-ilp-ass1}
\sum_{s=1}^{N}
x_{si} = 1
\qquad
\forall i= 1, \ldots, M\,,
\end{equation}
\begin{equation}\label{fpgatris-ilp-ass2}
\sum_{t=1}^{T}
y_{tij} = 1
\qquad
\forall i= 1, \ldots, M,\,j=1, \ldots, \ell_i\,.
\end{equation}

\medskip
\paragraph{Boundary Constraints} 
Next, we ensure that a request does not exceed the FPGA's boundary by
forcing all slot assignment variables that would cause an infeasible
placement to be zero.\\
\begin{equation}\label{fpgatris-ilp-boundary}
\forall i,j,\,
s = s_{\mathrm{low}}, \ldots, s_{\mathrm{up}}:
x_{si}= 0
\end{equation}  
where
\begin{eqnarray*}
s_{\mathrm{low}} & := & 
\begin{cases}N-r_{ij} + 2, & r_{ij} > 0 \cr 1, & r_{ij} < 0\end{cases}
\;\;\mbox{\rm and}\\
s_{\mathrm{up}} & := & 
\begin{cases} N, & r_{ij} > 0 \cr -r_{ij}-1,  & r_{ij} < 0\end{cases}\,.
\end{eqnarray*}

\medskip
\paragraph{Order Constraints}
Now, we ensure that the processing order is maintained; that is, 
request $(i,j)$ of $m_i$ is not scheduled before
request $(i,j-1)$ is finished.
For every $(i,j)$ there is exactly one $t$ such that $y_{tij}=1$.
Thus, summing up $t\cdot y_{tij}$ over $t$ for fixed $i$ and $j$
yields the time step where request $(i,j)$ is scheduled. This yields:
\begin{equation}\label{fpgatris-ilp-order}
\sum_{t=1}^{T}
t\,y_{tij}- 
\sum_{t=1}^{T}
t\,y_{tij-1}> 0
\qquad
\forall i,j>0\,.
\end{equation}

\begin{figure}[ht]
\centerline{\input{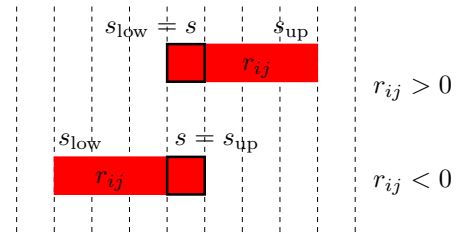}}
\caption{Occupancy Constraints: If $x_{si}=1$ and $y_{tij}=1$, 
the request $(i,j)$
occupies $r_{ij}$ slots left or right to $s$---depending on 
$\mathop{\rm sgn}(r_{ij})$.
\label{fpgatris-exclconstr-fig}}
\end{figure}

\medskip
\paragraph{Occupancy Constraints}
If $x_{si}=1$ and $y_{tij}=1$, 
the request $(i,j)$ occupies $r_{ij}$ slots adjacent to $s$ at time $t$;
see Fig.~\ref{fpgatris-exclconstr-fig}.
The first step to prevent other modules from overlapping with $m_i$ is to
set the appropriate occupancy variables as follows.\\
$\forall i = 1, \ldots, M,\, 
j = 1, \ldots, \ell_i,\, 
s = 1, \ldots, N,\,
t = 1, \ldots, T,\,$\\$
s' =  s_{\mathrm{low}}, \ldots, s_{\mathrm{up}}:$
\begin{equation}\label{fpgatris-ilp-occupancy}
x_{si} + y_{tij} - z_{s' tij}\leq 1\,,
\end{equation}
with 
\begin{eqnarray*}
s_{\mathrm{low}} & := &
\begin{cases}
s, & r_{ij} > 0 \cr
\max\{ 1, s+r_{ij} + 1\}, & r_{ij} < 0 
\end{cases}\;\;\mbox{\rm and}\\
s_{\mathrm{up}} & := &
\begin{cases}
\min\{ N, s+r_{ij} - 1\}, & r_{ij} > 0 \cr 
s, & r_{ij} < 0 
\end{cases}\,.
\end{eqnarray*}

\medskip
\paragraph{Exclusive Constraints}
By setting the appropriate occupancy variables with the occupancy constraints,
we can ensure that requests do not overlap. We 
allow at most one occupancy variable for a fixed slot and a fixed time
to be 1.\\
$\forall t = 1, \ldots, T,\, s = 1, \ldots, N:$
\begin{equation}\label{fpgatris-ilp-excl}
\sum_{i=1}^{M}\sum_{j=1}^{\ell_i} z_{stij}\leq 1
\end{equation}

\medskip
\paragraph{Delay Constraints}
If a request $(i, j+1)$ is delayed, the preceding request $(i,j)$ remains
on the FPGA until $(i, j+1)$ is scheduled.
Thus, if $z_{stij}=1$ either $z_{s(t+1)ij}$ must be set to 1---because
the module still occupies space on the FPGA---or 
$(i, j+1)$ is scheduled at time $t+1$; that is,
$y_{(t+1)i(j+1)}=1$ holds. The following constraints keep track of
delayed requests.\\
$\forall i = 1, \ldots M,\,
j = 1, \ldots, \ell_i-1,\,$\\$
s = 1, \ldots, N,\,
t = 1, \ldots, T-1:$
\begin{equation}\label{fpgatris-ilp-nopreemp}
z_{stij} - z_{s(t+1)ij} - y_{(t+1)i(j+1)}\leq 0
\end{equation}

\medskip
\paragraph{Usage Constraints}
Finally, we introduce some constraints that define our usage variables.
Let $u_t$ be 1, if at least one $y_{tij}$ is 1 or if 
$u_{t+1}$ is 1.\\
$\forall t = 1, \ldots T,\,
i = 1, \ldots, M,\, j = 1, \ldots, \ell_i:$
\begin{equation}\label{fpgatris-ilp-defineu1}
u_{t}- y_{tij}\geq 0
\end{equation}
and $\forall t = 2, \ldots, T:$
\begin{equation}\label{fpgatris-ilp-defineu2}
u_{t-1}- u_{t}\geq 0\,.
\end{equation}

\subsubsection{Objective Function}

To minimize the makespan, we use the following ILP:

\begin{eqnarray}\label{fpgatris-ilp-complete}
\min \raisebox{0pt}[13pt][0pt]{$\displaystyle \sum_{i=1}^{T}$} i\, u_{i}&
\mbox{\rm subject to} &
{\rm Eq.}~\ref{fpgatris-ilp-ass1}
\mbox{\rm --} \ref{fpgatris-ilp-defineu2}  \\
&& x_{si} \in \{0,1\} \nonumber \\
&& y_{tij} \in \{0,1\}\nonumber \\
&& z_{stij} \in \{0,1\}\nonumber \\
&& u_t \in \{0,1\}\,. \nonumber 
\end{eqnarray}

\subsection{Heuristic Methods}
We implemented several heuristics for our problem:
A simple FirstFit with and without delaying requests,
and two more elaborated heuristics, BestFit and TabuSearch.
Our methods pack the given modules in a semi-infinite strip.
The width of the strip is given by the number of slots on the FPGA; 
the height of
the strip corresponds to the time axis.

\subsubsection{FirstFit} Probably the simplest heuristic is to place the modules,
one by one, in a first-fit way into the strip: 
beginning with $s=1$ and $t=1$,
we test for every position if the module that must be placed overlaps with
already-placed modules. We choose the first position where no overlap 
occurs. Note that we disregard the possibility of delaying requests.

\pagebreak
~\vspace{-0.5cm}
\subsubsection{FirstFit with delays} 
This method works the same as the method above, but
allows the delaying of requests. That is, if for a certain start position 
the requests $0,\ldots,j-1$ fit into the strip without overlap, 
but request $j$ does not fit in time step $t'$, 
we search for the largest $j' < j$ such that
request $(i,j')$ fits in $t'$ and delay every request $j''=j'+1,\ldots$
(i.e., we move them upwards in the strip);
see Fig.~\ref{fpgatris-delay-fig}.

\subsubsection{BestFit} Similar to FirstFit with delays, we try to find a 
nonoverlapping position by testing every possible position. But now,
we do not choose the first feasible position, but we evaluate every
position as follows:
We separately count the unoccupied cells left and right to the placed module
and take the minimum of the these two values as a score for the given position.
For example, for the placement of $m_i$ in Fig.~\ref{fpgatris-delay-fig}
there are 4 unoccupied cells left to $m_i$ and 14 unoccupied cells
right to $m_i$, yielding a value of 4 for his placement.
We choose the position that yields the minimal score and break ties by
preferring the (first) position with least number of delays.

To avoid that every module is placed on the left or right side (yielding a
score of 0), we maintain an upper limit, $t_\mathrm{max}$, for the time.
Before we place a new module, $m_i$, we increase $t_\mathrm{max}$ by
$\ell_i/2$  and try to place $m_i$ within the given time bound. 
If this is not possible, we increase $t_\mathrm{max}$ by $\ell_i$ and
try again.

\begin{figure}[t]
\mbox{}\hrulefill\\
${\cal S} = (1,\ldots,M)$\\
{\bf for} $i = 0$ {\bf to} $M/2$\\
\mbox{}\quad ${\it found} = 0$\\
\mbox{}\quad {\bf for} $j = 1$ {\bf to} $M$\\
\mbox{}\qquad {\bf if} $(j, ((j + i) \bmod M)+1)$ are not in the tabu list\\
\mbox{}\qquad\quad Swap items at pos.\ $j$
and $((j + i) \bmod M)+1$ in ${\cal S}$\\
\mbox{}\qquad\quad Calculate makespan of BestFit\\
\mbox{}\qquad\quad {\bf if} makespan is the best so far {\bf then}\\ 
\mbox{}\qquad\qquad ${\it found} = j$\\
\mbox{}\qquad\quad Undo swapping\\
\mbox{}\quad {\bf if} ${\it found} > 0$  {\bf then}\\
\mbox{}\qquad Swap pos.\ ${\it found}$ and $(({\it found} + i) \bmod M)+1$
in ${\cal S}$\\
\mbox{}\qquad Store $({\it found}, (({\it found} + i) \bmod M)+1)$ 
in the tabu list\\
\mbox{}\hrulefill
\caption{Tabu Search\label{fpgatris:tabusearch}}
\end{figure}

\begin{figure*}[t]
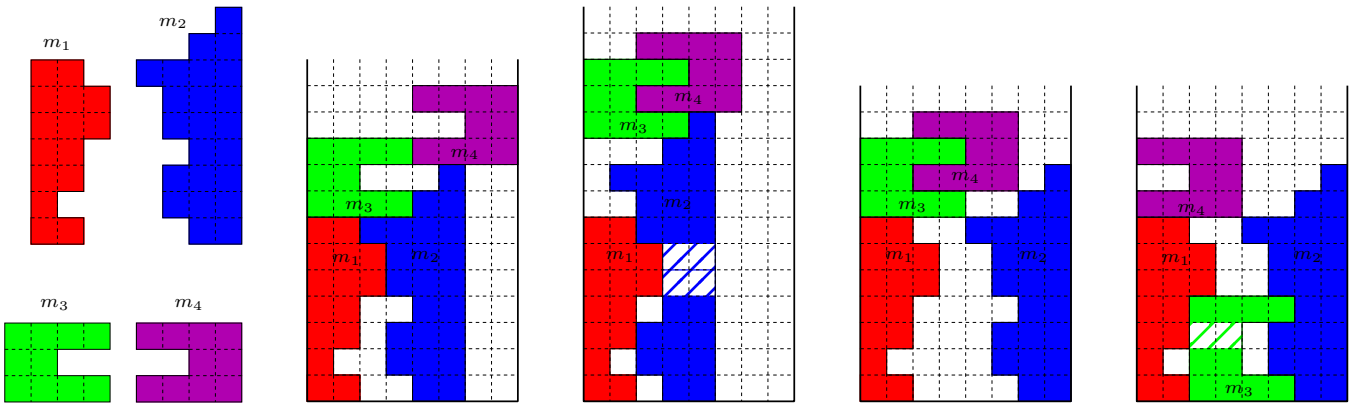

\centerline{\input{images/Example2.pstex_t} \qquad 
\input{images/ExampleFFnoDelay.pstex_t} \qquad \input{images/ExampleFF.pstex_t}
\qquad \input{images/ExampleBF.pstex_t} \qquad \input{images/ExampleTS.pstex_t}
}
\caption{From left to right: Example input and packings 
generated by (from left to right) FirstFit without delays,
FirstFit with delays, BestFit, and
TabuSearch/ILP. Delayed modules are shown hatched.
Note that $m_3$ is packed by BestFit on top of $m_1$, because this position
has value 0 and fits into the strip of height $t_\mathrm{max} + \ell_3/2$.
TabuSearch swaps $m_3$ and $m_4$ in the insertion order.
\label{fpgatris-packs-fig}}
\end{figure*}

\subsubsection{TabuSearch\label{tabu-sect}} 
BestFit inserts the modules in the given order (i.e., $m_1, m_2, \ldots, m_M$).
Obviously, the result of BestFit highly depends on the insertion order,
so we may get a better result if we permute the insertion order of the modules.
Thus, we use a tabu search to try several BestFit runs, each one
using a different order for the insertion of modules. 
Starting with the sequence ${\cal S} = (1,\ldots,M)$, we swap two
items of the sequence and compute the makespan that is achieved by
BestFit. More precisely, we maintain a {\em swapping distance}, $i$, ranging 
from $i=0$ to $M/2$.
For a fixed $i$,
we swap the items at positions $j$
and $((j+i) \bmod M)+1$ for $j=1,\ldots M$,
keeping track of the best makespan achieved so far. We accept the
swap that achieves the best makespan known so far.
A tabu list ensures that we do not swap an already accepted pair again;
see Fig.~\ref{fpgatris:tabusearch}.

\subsection{Experimental Results}
An example instance and solutions 
are shown in Fig.~\ref{fpgatris-packs-fig}.
The corresponding ILP was solved
in approximately 6 hours on an Intel(R) Xeon(TM) 3.20\,GHz CPU running
ILOG CPLEX 10.00 under Linux.
Note that TabuSearch yields the same result in less than one second.

To test our heuristics, we conducted a set of experiments.
For upper limits on the size of a request, 
$r_\mathrm{max}$,
ranging from 10\% to 90\% in steps of 10, we randomly generated sequences,
each of
20 modules. For each value of $r_\mathrm{max}$, we shuffled
20 sequences as follows: For every module, we choose its length, $\ell_i$,
randomly, by normal distribution with expected value 
$\mu = 10$ and variance $\sigma^2 = 5$.
The size for every request, $r_{ij}$, was chosen by 
normal distribution, too, with
an expected value of $\mu = r_\mathrm{max}/2$ and variance 
$\sigma^2 = r_\mathrm{max}/4$.
We present the results for $N = 50$, other FPGA widths showed similar 
results.

\begin{table}[h]
\centering
\begin{tabular}{|c|c|} \hline
Heuristic & Average running time\\ \hline
FirstFit & 0.25~s\\
FF with delays & 0.33~s\\
BestFit & 2.09~s\\
Tabu Search & 1125.06~s\\ \hline
\end{tabular}
\caption{Average running time for our heuristics.\label{results-tab}}
\end{table}

\begin{figure}[t]
\centerline{\psfig{figure=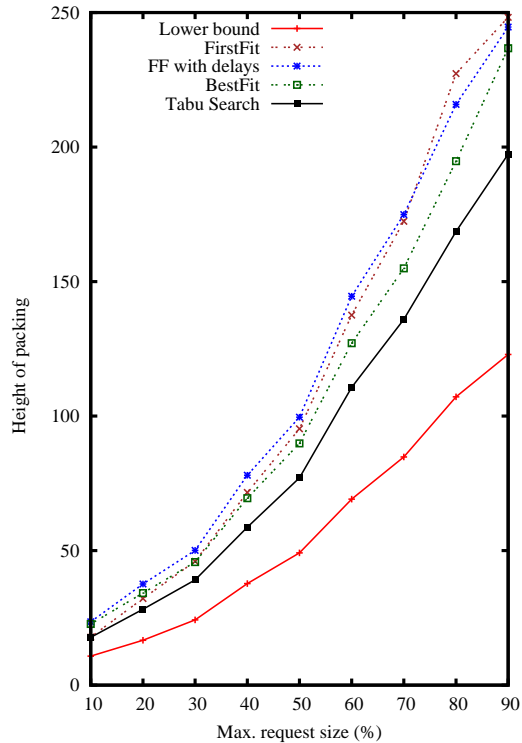,width=7cm}}
\caption{
A comparison of our 
heuristics---FirstFit (without delays),  FirstFit (with delays), BestFit, 
and TabuSearch---in settings with 
different densities (i.e., maximal value for a request) averaged over 100 runs
per densities, For comparison, a lower bound
(ratio of total area by number of slots) is shown.
\label{fpgatris-plot-fig}}
\end{figure}

\begin{figure}[t]
\centerline{\psfig{figure=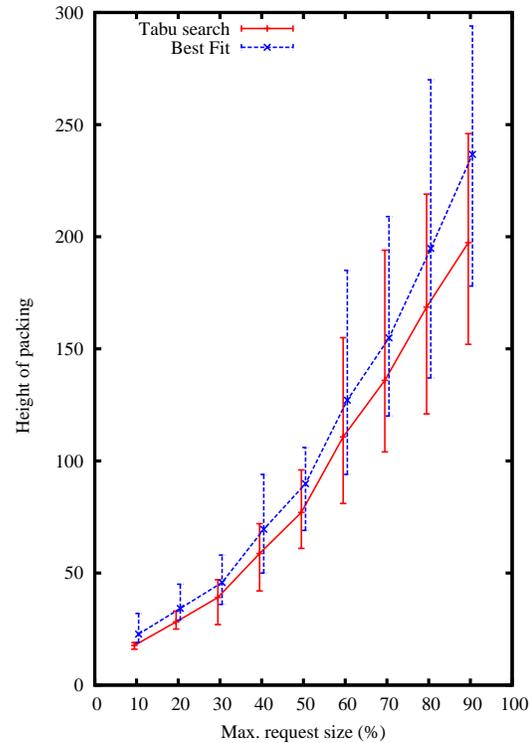,width=7cm}}
\caption{Mean-, maximal-, and minimal value averaged over 100 runs
per densities for BestFit and TabuSearch.\newline
~\newline
~\newline
\label{fpgatris-ploterr-fig}}
\end{figure}

Table~\ref{results-tab} 
shows the average running time over all experiments, 
Fig.~\ref{fpgatris-plot-fig} shows the mean value over 20 runs for
every heuristic and value of $r_\mathrm{max}$.
Fig.~\ref{fpgatris-ploterr-fig} shows mean-, maximal-, and minimal values
for BestFit and TabuSearch. 
For comparison, Fig.~\ref{fpgatris-plot-fig} also shows an 
average lower bound computed as the smallest area needed to
pack all requests; that is,
$$\mbox{LB} = \frac1N \sum_{i=1}^M \sum_{j=1}^{\ell_i} r_{ij}\,.$$
Choosing the best suited strategy depends on the scenario. For systems
that run the same request sequence on and on and that are produced in a
large number, it may even pay off to use the ILP. Clearly, balancing
computation time and quality, the tabu search is a better choice.
Nevertheless, it requires that the requests are known beforehand.
If this is not given, BestFit can be used, because it works in an
offline scenario as well as in an online setting.

\section{Conclusion}
Reconfigurable devices increasingly offer enough computing resources,
so that in the near future, multiple applications may be executed on
them concurrently, instead of just a single application. However,
until now there is a general lack of research on how to successfully
achieve secure and flexible execution of multiple applications on
dynamically partially reconfigurable devices.  We present an approach
called {\em virtual area management}, which is heavily influenced by
multitasking and operating system concepts of the software. Advantages
of our concept (e.g., support for accounting of resources, resource
protection, multiple scheduling and placement strategies, and a new
programming model) are explained. Furthermore, challenges posed by
this concept and possible solutions are discussed. Afterwards, a
specific approach to virtual area management in the offline case is
presented in more detail. It is based on the assumption that most
applications can handle also a delayed resource grant. The
corresponding optimization problem to minimize the total makespan is
solved with an ILP and heuristics. Both approaches are compared in an
experimental study.

The presented concept is a practicable solution to the considered
important problem; the proposed methods can be
applied to nowadays available reconfigurable devices. 
We do not rely on the assumption that saving and storing
hardware task states will no longer be considered as too expensive in the
future. Furthermore, programming models for reconfigurable
architectures, resource accounting and protection, bitstream
relocation and position independence, are all formidable research
problems on their own, however the concept is compatible and
extendable to different solution approaches to these problems.

\bibliographystyle{IEEEtran}


\end{document}